\documentclass[aps,prl,superscriptaddress,twocolumn,longbibliography]{revtex4-2}
\usepackage{bbm}
\usepackage{graphicx}% Include figure files
\usepackage{dcolumn}% Align table columns on decimal point
\usepackage{bm}% bold math
\usepackage{subfigure}
\usepackage{amsmath}
\usepackage{feynmf}
\usepackage{hyperref}

\usepackage{attachfile}

\newcommand{\bs}{\boldsymbol s}

\newcommand{\bk}{\boldsymbol k}

\newcommand{\bq}{\boldsymbol q}
\newcommand{\br}{\boldsymbol r}

\newcommand{\bR}{\boldsymbol R}

\usepackage{times}

\begin{document}

\title{Intrinsic Second-Order Topological Superconductors with Tunable Majorana Zero Modes}

\author{Xiao-Jiao Wang}
\affiliation{School of Physics, Sun Yat-sen University, Guangzhou 510275, China}
\affiliation{Guangdong Provincial Key Laboratory of Magnetoelectric Physics and Devices, Sun Yat-sen University, Guangzhou 510275, China}
\affiliation{State Key Laboratory of Optoelectronic Materials and Technologies, Sun Yat-sen University, Guangzhou 510275, China}

\author{Yijie Mo}
\affiliation{School of Physics, Sun Yat-sen University, Guangzhou 510275, China}
\affiliation{Guangdong Provincial Key Laboratory of Magnetoelectric Physics and Devices, Sun Yat-sen University, Guangzhou 510275, China}
\affiliation{State Key Laboratory of Optoelectronic Materials and Technologies, Sun Yat-sen University, Guangzhou 510275, China}

\author{Zhi Wang}
\affiliation{School of Physics, Sun Yat-sen University, Guangzhou 510275, China}
\affiliation{Guangdong Provincial Key Laboratory of Magnetoelectric Physics and Devices, Sun Yat-sen University, Guangzhou 510275, China}

\author{Zhigang Wu}
\email{wuzhigang@quantumsc.cn}
\affiliation{Quantum Science Center of Guangdong-Hong Kong-Macao Greater Bay Area (Guangdong), Shenzhen 508045, China}

\author{Zhongbo Yan}
\email{yanzhb5@mail.sysu.edu.cn}
\affiliation{School of Physics, Sun Yat-sen University, Guangzhou 510275, China}
\affiliation{Guangdong Provincial Key Laboratory of Magnetoelectric Physics and Devices, Sun Yat-sen University, Guangzhou 510275, China}
\affiliation{State Key Laboratory of Optoelectronic Materials and Technologies, Sun Yat-sen University, Guangzhou 510275, China}

\date{\today}

\begin{abstract}
Dirac semimetals, with their protected Dirac points, present an ideal platform 
for realizing intrinsic topological superconductivity. In this work, 
we investigate superconductivity in a two-dimensional, square-lattice nonsymmorphic Dirac semimetal. 
In the normal state near half-filling, the Fermi surface consists of two distinct pockets, each enclosing a Dirac point at a time-reversal invariant momentum ($\textbf{X}=(\pi,0)$ and $\textbf{Y}=(0,\pi)$). Considering an on-site repulsive and nearest-neighbor attractive
interaction, we use self-consistent mean-field theory to determine the ground-state pairing symmetry. 
We find that an even-parity, spin-singlet $d_{x^{2}-y^{2}}$-wave pairing is favored as it gives rise to 
a fully gapped superconducting state.
Since the pairing amplitude has opposite signs on the two Dirac Fermi pockets, 
the superconducting state is identified as a second-order topological superconductor. 
The hallmark of this topological phase is the emergence of Majorana zero modes at the system's boundaries. 
Notably, the positions of these Majorana modes are highly controllable and can be manipulated simply 
by tailoring the boundary sublattice terminations. Our results highlight the promise of 
nonsymmorphic Dirac semimetals for realizing and manipulating Majorana modes. 
\end{abstract}

\maketitle

Topological superconductors (TSCs)  are highly sought-after materials
for hosting Majorana zero modes (MZMs)~\cite{qi2011,alicea2012new,leijnse2012introduction,Tanaka2012,stanescu2013majorana,
Beenakker2013,Elliott2015,Sato2016jpsj}. The non-Abelian statistics of these quasiparticles
offer a promising route to fault-tolerant topological quantum computation~\cite{nayak2008review,sarma2015majorana,Karzig2017MZM,Marra2022}.
Over the past two decades, guided by pioneering theoretical work~\cite{read2000,kitaev2001unpaired,fu2008,sato2009non,lutchyn2010,oreg2010helical,alicea2010}, 
significant progress has been made in engineering topological 
superconducting phases, principally by proximitizing 
topological insulators~\cite{Xu2015MZM,Sun2016Majorana,Liu2024vortex}, magnetic chains~\cite{Nadj2014MZM,Jeon2017} or semiconducting nanowires~\cite{Mourik2012MZM,das2012zero,deng2012anomalous,finck2013} with conventional $s$-wave 
superconductors. However, these heterostructure approaches 
currently face significant challenges~\cite{Prada2020,Flensberg2021,Sarma2021,Yazdani2023}, including 
impurity-induced degradation of sample quality, small superconducting gaps, and the presence of trivial 
bound states whose experimental signatures can resemble those of MZMs.  The discovery of intrinsic TSCs--those that do not rely on proximity effects--offers a route to mitigating these issues, as such systems generally promise larger superconducting gaps and lower levels of disorder.

Topological materials that become superconducting below a critical temperature 
are promising candidates for intrinsic TSCs. A  prominent example is provided by certain three-dimensional 
iron-based superconductors which feature topologically inverted band structures and Dirac surface states~\cite{Wang2015iron,Xu2016FeSC,zhang2018iron,Zhang2019hinge,Majid2022}. 
In these materials, compelling evidence for MZMs has been observed at vortex line ends~\cite{wang2018evidence,Liu2018MZM,machida2019zero,kong2019half}.
Despite this progress, manipulating MZMs in such materials remains a major experimental challenge, 
underscoring the need for intrinsic TSCs capable of hosting highly tunable Majorana modes.

\begin{figure}[t]
\centering
\includegraphics[width=0.45\textwidth]{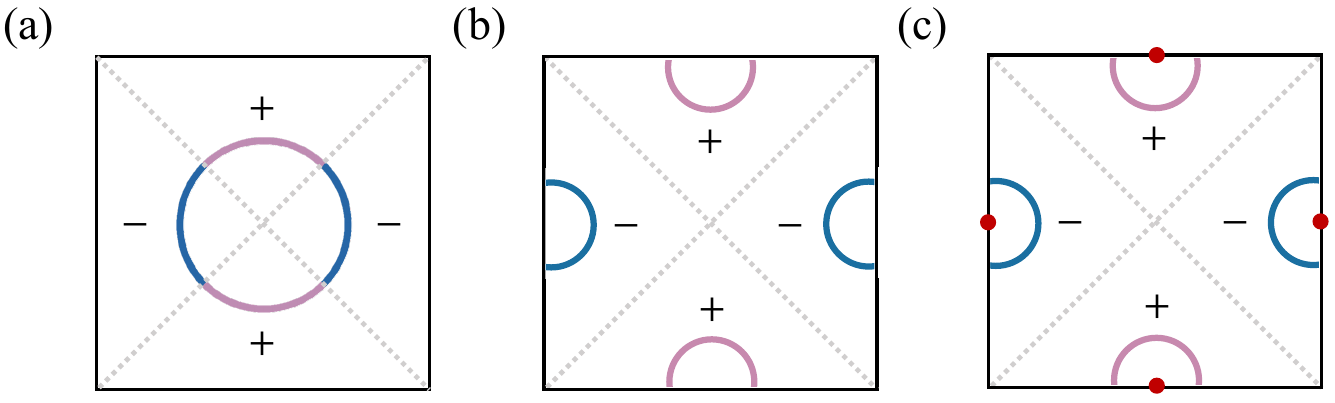}
\caption{Illustration of how the Fermi surface (solid lines) 
dictates superconducting properties for a $d$-wave pairing state. 
The dashed lines denote the pairing nodes. (a) Nodal phase.
(b) Fully gapped phase with trivial topology.
(c) Fully gapped phase with nontrivial topology, 
where the Fermi pockets enclose a Dirac point (red dots).
}\label{fig1}
\end{figure}

In the search for intrinsic TSCs, the Fermi surface and pairing symmetry are two key factors to consider~\cite{qi2010d,sato2010odd,fu2010odd}.
While odd-parity spin-triplet pairings or phase-winding pairings (e.g., $d\pm id$) naturally give rise to topological superconductivity~\cite{Qi2009b,Schaffer2012CTSC,Liu2013CTSC,Qin2019CTSC,scammell2022intrinsic,Jahin2023HOTSC}, 
materials that host such exotic pairing states are 
exceedingly rare and typically have very low transition temperatures ($T_c$)~\cite{Saxena2000,Aoki2001,Huy2007,Ran2019}. 
In contrast, the most common pairing symmetries found in high-$T_c$ materials, 
like the even-parity spin-singlet $d$-wave and extended $s$-wave observed in cuprates 
and iron-based systems~\cite{Harlingen1995,Tsuei2000,Damascelli2003,Paglione2010,Wang2011iron,Hirschfeld2011,Chubukov2012}, generally 
lead to nodal [Fig.~\ref{fig1}(a)] or topologically trivial gapped states [Fig.~\ref{fig1}(b)].
To achieve topological superconducting states from these even-parity, spin-singlet pairings with a fixed phase, 
the Fermi surface needs to enclose a band degeneracy so that it acquires
nontrivial spin- or orbital-momentum locking~\cite{zhang2013kramers,Hosur2014}. Furthermore, the pairing must change sign 
between different Fermi pockets [Fig.~\ref{fig1}(c)], locking the nodes and the Fermi surface 
into a stable topological configuration that protects against 
adiabatic deformation to the trivial limit~\cite{Yan2020configuration}.

In this work, we explore intrinsic topological superconductivity in a two-dimensional, 
square-lattice nonsymmorphic Dirac semimetal (DSM) possessing $\mathcal{PT}$ symmetry.
The band structure features Dirac points at the time-reversal invariant momentums (TRIMs)
$\textbf{X}=(\pi,0)$ and $\textbf{Y}=(0,\pi)$, yielding two distinct Fermi pockets
near half-filling, each enclosing one Dirac point. Using self-consistent mean-field 
theory for an on-site repulsive and nearest-neighbor attractive interaction, 
we find that the superconducting ground state favors an even-parity, spin-singlet $d_{x^{2}-y^{2}}$-wave pairing.
This pairing, which changes sign between the two pockets, results in a fully gapped phase identified 
as a second-order TSC. Notably, the resulting MZMs are highly tunable; unlike their typical pinning 
at sharp corners, their location along the edge can be engineered at will by controlling the boundary 
sublattice terminations.

\begin{figure}[t]
\centering
\includegraphics[width=0.45\textwidth]{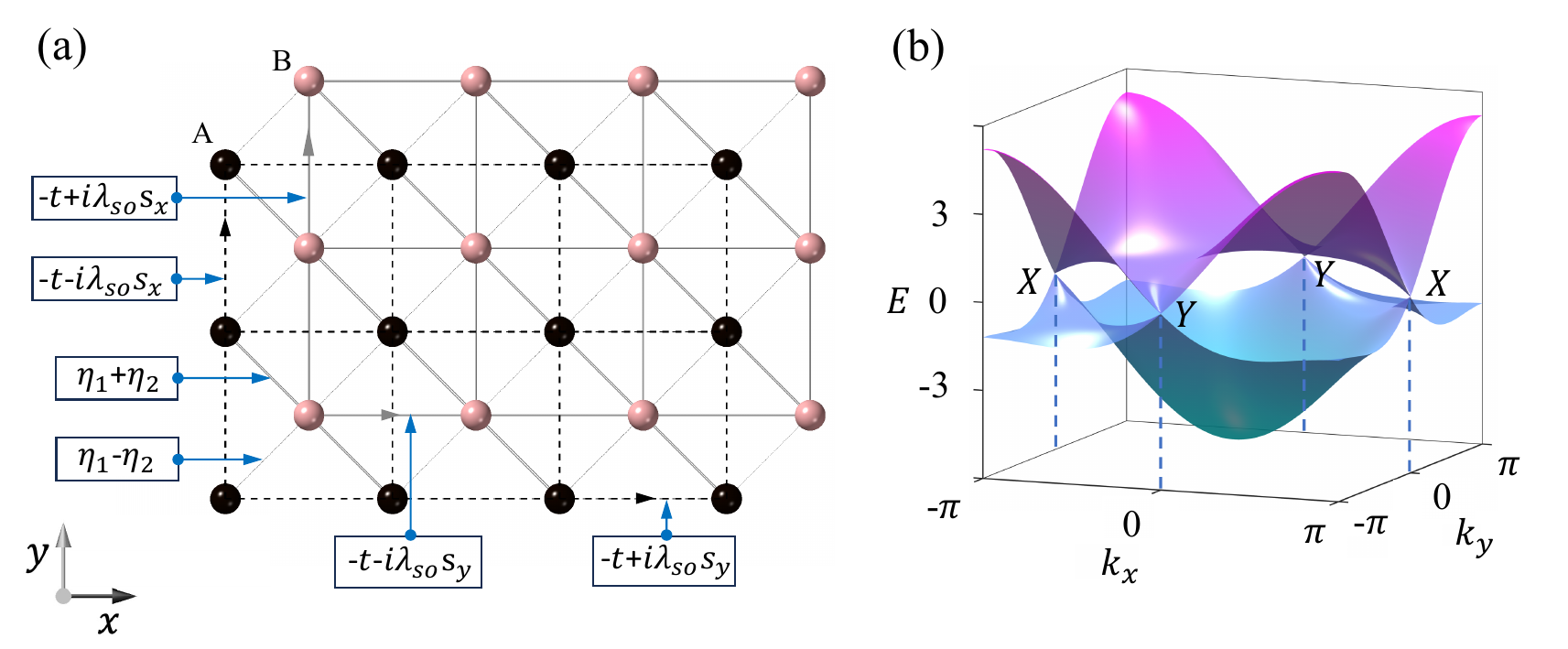}
\caption{ (a) Sketch of a top-down view of the bilayer lattice,
where the two layers are shifted by a vector $(1/2,1/2)a$. The hopping and spin-orbit 
coupling coefficients are shown.  (b) The band structure of the DSM, where  
the Dirac points appear at $\textbf{X}= (\pi,0)$, $\textbf{Y}= (0,\pi)$. 
Parameters are $\{t, \lambda_{\rm so}, \eta_1, \eta_2\} = \{0.5, 0.4, 0.8, 0.8\}$.
}\label{fig2}
\end{figure}

{\it DSM Hamiltonian.---}We consider a DSM which consists of two layers 
that are relatively shifted by a vector $(1/2,1/2)a$, where 
$a$ denotes the lattice constant. 
From a top-down view, the bilayer forms a bipartite square lattice with primitive lattice 
vectors ${\bm a}_1 = a \hat x$ and ${\bm a}_2 = a \hat y$, as illustrated in Fig.~\ref{fig2}(a).
The corresponding tight-binding Hamiltonian is given by
\begin{align}
H_{0}=&-\sum_{\br,j,\alpha}\left \{c_{\br,\alpha}^{\dag}[t+i\lambda_{\rm so}(\hat{{\bm a}}_j\times \bs)_{z}\xi_{\alpha}]c_{\br+{\bm a}_j,\alpha} +h.c. \right\}\nonumber \\
&+\sum_{\br, \beta}[\eta_{\beta}(c_{\br,A}^{\dag}c_{\br+ {\bm \delta}_\beta,B}+c_{\br,A}^{\dag}c_{\br- {\bm \delta}_\beta,B})+h.c.],
\end{align}
where $\br$ denotes the Bravais lattice sites of either layer, specified by $\alpha=\{A,B\}$, $\hat{{\bm a}}_j={\bm a}_{j}/a$ represents  
the unit vector along the $j$ direction, and $c_{\br,\alpha}^{\dag}\equiv(c_{\br,\alpha\uparrow}^{\dag},c_{\br,\alpha\downarrow}^{\dag})$  
creates a fermion with spin $s$ on the lattice $\br$ of the layer $\alpha$. The first line of 
the Hamiltonian describes the intra-layer nearest-neighbor hopping with amplitude $t$ and the 
Rashba spin-orbit coupling of strength $\lambda_{\rm so}$. The parameter $\xi_{\alpha}$ equals $+1$ for the top layer (A) and $-1$
for the bottom layer (B), indicating that the spin polarizations induced by the spin–orbit coupling are opposite in the two layers. The second line describes the interlayer nearest-neighbor hoppings along the two diagonal directions ${\bm \delta}_{\pm}= \frac{1}{2}(-{\bm a}_1\pm {\bm a}_2) $, with corresponding amplitudes $\eta_{\pm}=\eta_{1}\pm\eta_{2}$ [see Fig.~\ref{fig2}(a)]. It is worth pointing out that the particular form of spin-orbit coupling arises in systems where inversion symmetry is locally broken but globally preserved~\cite{Zhang2014hidden}. A representative example 
is the iron-based superconductor FeSe~\cite{Qin2022hosc,Zhang2024TSC}. Although the crystal as a whole is 
inversion-symmetric, the unit cell contains two inequivalent Fe atoms, each of which is asymmetrically 
coordinated by surrounding Se atoms, leading to a local breaking of inversion symmetry.

By performing a Fourier transformation from the real space to the momentum space, 
we obtain $H_{0}=\sum_{\bk}\psi_{\bk}^{\dag}\mathcal{H}(\bk)\psi_{\bk}$, where the basis 
function is $\psi_{\bk}^{\dag}=(c_{\bk,A\uparrow}^{\dag},c_{\bk,A\downarrow}^{\dag},c_{\bk,B\uparrow}^{\dag},c_{\bk,B\downarrow}^{\dag})$
and the momentum-space Hamiltonian is 
\begin{eqnarray}
\mathcal{H}(\bk)=\epsilon(\bk)+\eta(\bk)\sigma_x+2 \lambda_{\rm so}\sigma_z(\sin k_xs_y-\sin k_y s_x).
\end{eqnarray}
Here we define the functions $\epsilon(\bk) = -2t(\cos k_x + \cos k_y)$ and 
$\eta(\bk) = 4[\eta_1 \cos (k_x/2) \cos (k_y/2) + \eta_2 \sin (k_x/2) \sin (k_y/2)]$ for brevity. 
The Pauli matrices  $\sigma_{i}$ and $s_{i}$ act on the layer ($A,B$) and spin ($\uparrow,\downarrow$) space, respectively. For notational simplicity, we set the lattice constant $a$ to unit and omit all identity matrices throughout 
the paper. This Hamiltonian possesses $\mathcal{PT}$ symmetry ($\mathcal{PT}=\sigma_{x}s_{y}\mathcal{K}$ 
with $\mathcal{K}$ the complex conjugation operator), and a glide symmetry 
($\{\mathcal{M}_{z}|(\frac{1}{2},\frac{1}{2})\}=i\sigma_{x}s_{z}$). The latter symmetry operation consists of  a mirror reflection 
about the middle plane of the two layers  $\mathcal{M}_{z}$ followed by a translation denoted by $(\frac{1}{2},\frac{1}{2})$.
When $\eta_{2}=0$, the Hamiltonian further possesses two screw symmetries, 
$\{\mathcal{C}_{2x}|(\frac{1}{2},0)\}=i\sigma_{x}s_{x}$ and $\{\mathcal{C}_{2y}|(0,\frac{1}{2})\}=i\sigma_{x}s_{y}$, where 
$\mathcal{C}_{2a}$ denotes a $\pi$ rotation about the $a$-axis~\cite{Young2015DSM,Mo2025a}.

The coexistence of $\mathcal{PT}$ symmetry and nonsymmophic symmetry admits robust 
Dirac points in the band structure~\cite{Young2015DSM}. When $\eta_{2}=0$, there are three  Dirac points, 
located at $\textbf{X}= (\pi,0)$, $\textbf{Y}= (0,\pi)$ and $\textbf{M} = (\pi,\pi)$. 
Once $\eta_{2}$ becomes finite, the Dirac point at $\textbf{M}$ is gapped, due to the breaking 
of the two screw symmetries. However, the two Dirac points at $\textbf{X}$ and $\textbf{Y}$ remain 
intact, as shown in Fig.~\ref{fig2}(b). We focus on this general case which admits two well-separated, disconnected Fermi pockets, 
each enclosing a Dirac point. 

The presence of Dirac points, which are momentum-space topological defects with singular quantum geometry, 
renders these Fermi pockets fundamentally distinct from conventional Fermi pockets that do not enclose them. 
We refer to the former as Dirac Fermi pockets (DFPs) and the latter normal Fermi 
pockets (NFPs). This distinction is underscored by a key topological constraint: a NFP 
can be continuously shrunk to a point and vanish under variation of system parameters 
(e.g., the chemical potential) without breaking any symmetry. In contrast, a DFP is 
topologically protected; it cannot vanish on its own and must instead annihilate in pairs with another DFP~\cite{Mo2025b}. 
This inherent topological stability makes investigating the pairing symmetry in this system 
a subject of fundamental interest.

{\it Pairing symmetry and spectrum.---}We first classify the pairing channels based on 
symmetry. The momentum-space Hamiltonian possesses three crystal symmetries: 
inversion symmetry ($\mathcal{P}\mathcal{H}(\bk)\mathcal{P}^{-1}=\mathcal{H}(-\bk)$ with $\mathcal{P}=\sigma_{x}$), 
in-plane mirror symmetry ($\mathcal{M}_{z}\mathcal{H}(\bk)\mathcal{M}_{z}^{-1}=\mathcal{H}(\bk)$ with $\mathcal{M}_{z}=i\sigma_{x}s_{z}$), and $\mathcal{C}_{2z}$ rotation symmetry ($\mathcal{C}_{2z}\mathcal{H}(\bk)\mathcal{C}_{2z}^{-1}=\mathcal{H}(-\bk)$ with $\mathcal{C}_{2z}=is_{z}$).  Therefore, the pairing symmetry can be classified by the $\mathcal{C}_{2h}$ group, which has four one-dimensional 
irreducible representations (IRs): two even-parity channels ($A_g$, $B_g$) and two odd-parity 
channels ($A_u$, $B_u$). Pairing in the $A_g$ channel is conventionally classified as $s$-wave 
pairing because its order parameter is invariant under all symmetry operations of the group. 
However, its corresponding order parameter is not necessarily a constant.

Before considering specific interactions, we first outline the key 
characteristics of the superconducting states for the four IRs, 
including their energy spectra and topological properties. 
Since the pairings in these one-dimensional IRs preserve time-reversal symmetry, 
the resulting superconducting states all belong to symmetry class DIII~\cite{Schnyder2008,kitaev2009periodic,Ryu2010}. Within 
this class, the first-order topology is characterized by a $Z_{2}$ invariant.
As the inversion symmetry is also present, the $Z_{2}$ invariant can simply be 
defined as the product of parity eigenvalues of the negative-energy bands~\cite{fu2007a}. 
To facilitate a concrete analysis, we restrict the real-space pairing to nearest neighbors. 
This allows us to explicitly write down the general pairing function for each IR. In the Nambu basis,
defined as $\Psi_{\bk}^{\dag}=(\psi_{\bk}^{\dag},\psi_{-\bk}^{T})$, 
the resulting pairing terms are as follows:

\begin{figure}[t]
\centering
\includegraphics[width=0.45\textwidth]{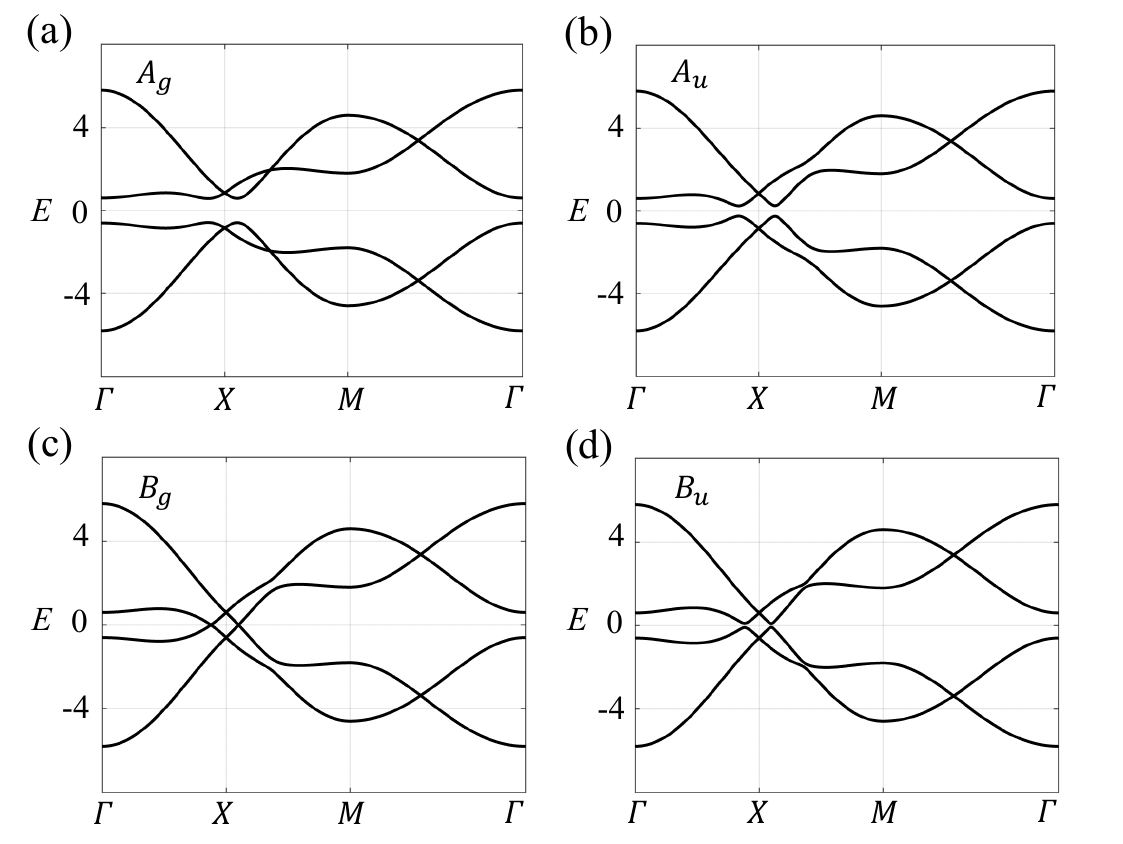}
\caption{BdG spectra corresponding to the pairing functions in the four different IRs. 
All spectra are calculated with the common parameter set: 
$\{t, \lambda_{\rm so}, \eta_1, \eta_2, \mu, \Delta_{1x},\Delta_{1y}\} = \{0.5, 0.4, 0.8, 0.8, 0.6, 0.3,-0.3\}$. 
For $A_{g}$ and $A_{u}$, $\Delta_{0}$ is set to zero. 
For $B_{g}$ and $B_{u}$, we choose $M_{\Delta}=\tau_{y}\sigma_{z}s_{x}$ and $M_{\Delta}^{\prime}=\tau_{y}s_{x}$, respectively. 
}\label{fig3}
\end{figure}

{\it $A_{g}$}: $(\Delta_{0}+\Delta_{1x}\cos k_{x}+\Delta_{1y}\cos k_{y})\tau_{y}s_{y}$, 
where $\tau_{i}$ are the Pauli matrices in particle-hole space. This even-parity, spin-singlet pairing leads a 
fully-gapped spectrum as long as the pairing nodes, if existing, do not cross the Fermi surface.
In the gapped regime, the $Z_{2}$ invariant is always trivial, indicating that this pairing cannot give rise to  
first-order TSCs. However, as we will demonstrate later, this pairing can result in a 
second-order TSC, provided that the pairing nodes 
and the Fermi surface form a configuration similar to that in Fig.~\ref{fig1}(c).

{\it $B_{g}$}: $(\Delta_{1x}\sin k_{x}+\Delta_{1y}\sin k_{y})M_{\Delta}$, where $M_{\Delta}=\{\tau_{y}\sigma_{z}s_{x},\tau_{y}\sigma_{z}s_{z}, \tau_{x}\sigma_{z}\}$. This channel corresponds to spin-triplet pairings. 
Because the pairing function vanishes identically at all TRIMs, 
the spectrum will always exhibit nodes. Therefore, this pairing channel gives rise to 
nodal superconducting phases.  

{\it $A_{u}$}$: (\Delta_{0}+\Delta_{1x}\cos k_{x}+\Delta_{1y}\cos k_{y})\tau_{y}\sigma_{z}s_{y}$. 
This odd-parity, spin-singlet pairing produces a gapped spectrum unless a pairing node crosses the Fermi surface.
Unlike the $A_{g}$ case, this pairing can stabilize a first-order TSC phase. In the weak-pairing limit, the 
$Z_{2}$ invariant simplifies to a criterion based on the even or odd number of Fermi surfaces enclosing TRIMs~\cite{sato2010odd,fu2010odd}. 
When there is a single Fermi surface, the $Z_{2}$ invariant is nontrivial, leading to a first-order TSC 
characterized by a pair of helical Majorana edge states~\cite{Mo2025b}.

{\it $B_{u}$}: $(\Delta_{1x}\sin k_{x}+\Delta_{1y}\sin k_{y})M_{\Delta}^{\prime}$, 
where $M_{\Delta}^{\prime}=\{\tau_{y}s_{x},\tau_{y}s_{z}, \tau_{x}\}$. 
Similar to the $B_{g}$ case, the superconducting state always hosts a nodal spectrum. 
Therefore, this pairing channel also leads to nodal superconducting phases.

To qualitatively assess which pairing channel is energetically  most stable, we compare the 
superconducting spectra for the four IRs under identical parameter conditions. Specifically, 
we consider a Fermi surface composed of two DFPs enclosing \textbf{X} and 
\textbf{Y}, respectively, and we set $\Delta_{0}=0$ and $\Delta_{1x}=-\Delta_{1y}$, as suggested by the solutions of the gap equation (see the following section). The calculated spectra along high-symmetry paths are shown in Fig.~\ref{fig3}. 
Consistent with our theoretical analysis, the $B_{g}$ and $B_{u}$ pairings
lead to a nodal spectrum, whereas the $A_{g}$ and $A_{u}$ pairings
lead to a fully-gapped spectrum.  Furthermore, the superconducting gap opened by the 
$A_{g}$ pairing is  noticeably larger than that of the $A_{u}$ pairing. These spectral features suggest that the 
$A_{g}$  channel is likely the leading pairing instability, a conclusion we will subsequently 
reinforce by analyzing a specific interaction.

{\it Interaction and favored pairing.---}We consider a short-range interaction:
\begin{eqnarray}
H_{int}=U\sum_{i,\alpha}n_{i,\alpha\uparrow}n_{i,\alpha \downarrow}+V\sum_{<ij>,\alpha}n_{i,\alpha\uparrow}n_{j,\alpha\downarrow},
\end{eqnarray}
where $U>0$ is the strength of the on-site repulsive Hubbard interaction, 
and $V<0$ is the strength of the attractive nearest-neighbor interaction 
within each layer ($\alpha=\{A,B\}$). The on-site repulsion $U$ originates 
from the Coulomb interaction, while the attractive $V$ can arise from 
electron-phonon coupling~\cite{Chen2021cuprate,Peng2023hubbard} or other bosonic fluctuations.
Here, we restrict the attractive nearest-neighbor interaction 
to electrons with opposite spin. While this interaction directly rules out spin-triplet pairings 
with total $S_{z}=\pm1$, it still admits the one with $S_{z}=0$. 
Therefore, this interaction is sufficient to generate pairing in all channels.

In momentum space the interaction reads
\begin{eqnarray}
H_{int}=\frac{1}{N}\sum_{\bk\bk'\bq,\alpha}V(\bq)c_{\bk+\bq, \alpha \uparrow}^{\dag}c_{\bk'-\bq, \alpha \downarrow}^{\dag}c_{\bk', \alpha \downarrow}c_{\bk, \alpha\uparrow},
\end{eqnarray}
where  $V(\bq)=U+2V(\cos q_{x}+\cos q_{y})$. 
We focus on zero-momentum pairings, for which the interaction can be simplified as 
\begin{eqnarray}
H_{int}\simeq\frac{1}{N}\sum_{\bk,\bk',\alpha}V(\bk-\bk')c_{\bk, \alpha\uparrow}^{\dag}c_{-\bk, \alpha\downarrow}^{\dag}c_{-\bk', \alpha\downarrow}c_{\bk', \alpha\uparrow}.
\end{eqnarray}

\begin{figure}[t]
\centering
\includegraphics[width=0.45\textwidth]{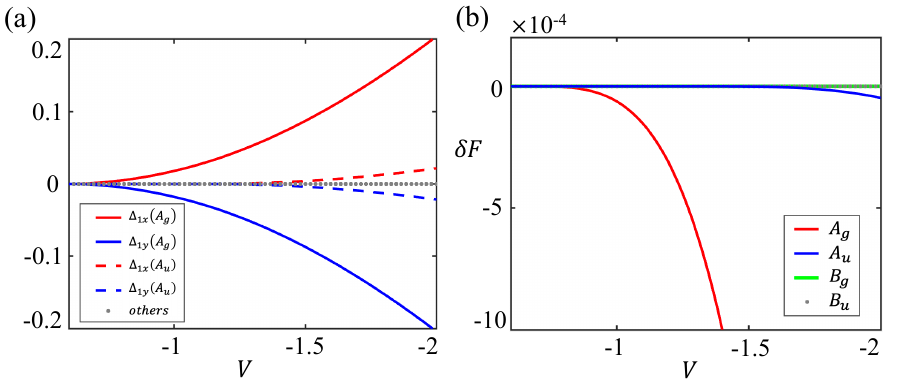}
\caption{(a) Pairing amplitudes for  the gap functions in the four IRs. (b) 
Free-energy difference $\delta F$ between the normal and 
superconducting states at zero temperature. Parameters 
are $\{t, \lambda_{\rm so}, \eta_1, \eta_2, \mu, U\} = \{0.5, 0.4, 0.8, 0.8, 0.6, 0.1\}$. 
}\label{fig4}
\end{figure}

\begin{figure*}[t]
\centering
\includegraphics[width=0.9\textwidth]{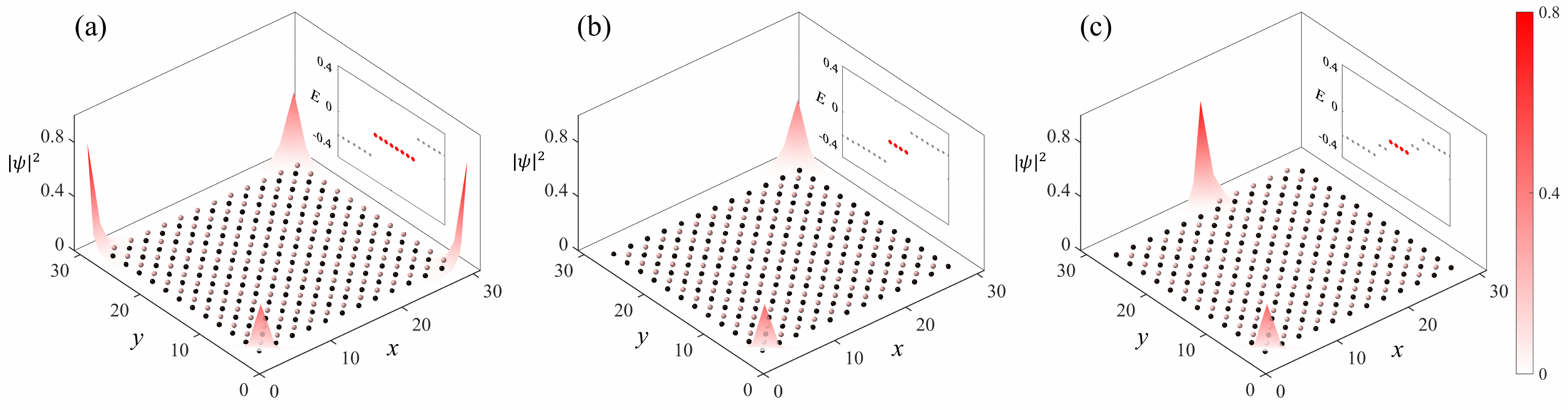}
\caption{Tunability of MZMs via sublattice termination. (a) With complete unit cells at all boundaries, 
the system hosts one Majorana Kramers pair at each corner. (b) Changing the sublattice termination on the 
upper ($y$-normal) and right ($x$-normal) edges halves the number of Majorana Kramers pairs. (c) Further 
adjusting the termination of the upper edge creates a domain wall and relocates a Majorana Kramers pair from the corner to 
a generic boundary site. The MZM count is quantified by the number of zero-energy states shown in the insets. 
Common parameters:  $\{t, \lambda_{\rm so}, \eta_1, \eta_2, \mu, \Delta\} = \{0.5, 0.4, 0.8, 0.8, 0.6, 0.3\}$
}\label{fig5}
\end{figure*}

Following the standard Bardeen-Cooper-Schrieffer (BCS) theory, we define the gap function as~\cite{Zhu2023TSC} 
\begin{eqnarray}
\Xi_{\alpha s,\alpha\bar{s}}(\bk)=-\frac{1}{N}\sum_{\bk'}V(\bk-\bk')\langle c_{\bk',\alpha s}c_{-\bk',\alpha\bar{s}}\rangle.
\end{eqnarray}
 Since pairing occurs only within each layer, the gap function $\Xi(\bk)$ as a matrix is diagonal in the layer index.
The Bogoliubov-de Gennes (BdG) mean-field Hamiltonian describing the superconducting state 
can be expressed as $H=\frac{1}{2}\sum_{\boldsymbol{k}}\Psi_{\boldsymbol{k}}^\dagger 
\mathcal H_{\text{BdG}}(\boldsymbol{k})\Psi_{\boldsymbol{k}}$, with
\begin{eqnarray}
		\mathcal H_{\text{BdG}}(\boldsymbol k)=\left(\begin{array}{cc}\mathcal H(\boldsymbol{k})-\mu & \Xi(\boldsymbol{k}) \\ \Xi^{\dagger}(\boldsymbol{k}) & -\mathcal H^*(-\boldsymbol{k})+\mu\end{array}\right). \label{eq4}
\end{eqnarray}
Here, $\mu$ is the chemical potential, and $ \Xi(\bk)$ is the pairing matrix. 
We employ a self-consistent procedure to determine the zero-temperature
pairing amplitudes for each IR, iterating until convergence is reached. 
In our calculations, we set $\mu$ to a value  where the Fermi surface 
contains two DFPs, fix the on-site repulsion $U$, and vary the nearest-neighbor 
attraction $V$ to track the evolution of the pairing. As expected from the on-site repulsion, 
the numerical results show a vanishingly small on-site spin-singlet pairing amplitude $\Delta_0$.
Furthermore, we find that $\Delta_{1x} = -\Delta_{1y}$ for all four pairing channels.  
Over the considered range of $V$, the $B_g$ and $B_u$ pairing amplitudes are negligible, 
while those of the $A_u$ channel becomes notable for $V< -1.5$. In contrast, the $A_g$ pairing amplitude 
is significantly larger than all others under the same conditions, as shown in Fig.~\ref{fig4}(a).
To identify the dominant superconducting channel, we compute the free energy difference $\delta F = F_S - F_N$ 
between the superconducting and normal states. The results in Fig.~\ref{fig4}(b) clearly indicate that the $A_g$ 
pairing channel is energetically favored, confirming it as the ground state.

{\it Second-order TSC with tunable MZMs.---} In this channel, the numerical results in Fig.~\ref{fig4}(a) 
show that $\Delta_0 =0$, and $\Delta_{1x}$ and $\Delta_{1y}$ take opposite values, resulting in a gap 
function of the form $\Xi(\bk) = -i\Delta(\cos k_{x} - \cos k_{y})s_{y}$ where $\Delta \equiv |\Delta_{1x}|$.
While this functional form is characteristic of a $d_{x^{2}-y^{2}}$-wave pairing, 
the $C_{2h}$ symmetry group of the system does not distinguish it from a 
conventional $s$-wave pairing, both belonging to the $A_g$ IR.
This is evidenced by the gap function having a fixed sign on each DFP, 
similar to the behavior of the gap function in a conventional $s$-wave superconductor. 
The key physical distinction, however, is that this gap function has opposite signs 
on the two DFPs located near  the $\textbf{X}$ and $\textbf{Y}$ points. As we demonstrate 
below, this sign structure directly leads to a second-order TSC with tunable MZMs.

The BdG Hamiltonian commutes with the mirror symmetry operator, $[\mathcal{H}_{\rm BdG}(\boldsymbol{k}), \tilde{\mathcal{M}}_{z}]=0$, where $\tilde{\mathcal{M}}_{z}= i\tau_{z}\sigma_{x}s_{z}$. Since $\tilde{\mathcal{M}}_{z}$ has eigenvalues $\pm i$, the Hamiltonian can be block-diagonalized as $\mathcal{H}_{\rm BdG} = \mathcal{H}_{+i} \oplus \mathcal{H}_{-i}$, corresponding to the two mirror subsectors. In the basis where $\tilde{\mathcal{M}}_{z}$ is diagonal, we have
\begin{eqnarray}
\mathcal{H}_{\pm i}(\bk)&=&(\epsilon(\bk)-\mu)\tau_{z}+\eta(\bk)\tau_{z}\rho_{z}\mp \Delta(\bk)\tau_{x}\rho_{z}\nonumber\\
&&+2\lambda_{\rm so}(\mp\sin k_{x}\rho_{y}+\sin k_{y}\rho_{x}),
\end{eqnarray}
where we define $\Delta(\bk)=\Delta(\cos k_{x}-\cos k_{y})$, and $\rho_{x,y,z}$ are Pauli matrices acting 
on the two-dimensional Hilbert space spanned by the eigenvectors of $\sigma_{x}s_{z}$ with positive or negative 
eigenvalues. 
Each mirror sector of the Hamiltonian possesses chiral symmetry, 
represented by the operator $S=\tau_{y}\rho_{z}$, and therefore belongs to the AIII class. 
This symmetry class does not support first-order topological gapped phases in two dimensions~\cite{Schnyder2008,kitaev2009periodic,Ryu2010}. 
Nevertheless, we find that each sector is characterized by a nontrivial quantized quadrupole moment 
$q_{xy}=1/2$~\cite{Benalcazar2017,Benalcazar2017prb,Kang2019multipole,Yang2021HOTI,Li2020quadrupole}, indicating a second-order 
topological phase. Further confirmation comes from diagonalizing 
the full Hamiltonian under open boundary conditions along both the $x$ and $y$ directions. Consistent with 
the bulk topological invariant, our computation reveals the presence of eight MZMs, localized as a Kramers pair 
at each corner, as shown in Fig.~\ref{fig5}(a). This explicitly demonstrates the system as a second-order TSC.

Remarkably, unlike typical second-order TSCs where MZMs are pinned at sharp corners~\cite{Langbehn2017,Geier2018,Khalaf2018,Zhu2018hosc,Yan2018hosc,
Wang2018weak,Wang2018hosc,Liu2018hosc,Hsu2018hosc,Wu2019hosc,Yan2019hosca,Yan2019hoscb,Volpez2019SOTSC,Zhang2019hoscb,Pan2019SOTSC,
Zhu2019mixed,Ahn2020hosc,Hsu2020hosc,Wu2020SOTSC,Majid2020hoscb,Franca2019SOTSC,
wu2020boundaryobstructedb,Laubscher2020mcm,Roy2020HOTSC,Li2021bts,Luo2021hosc,Ghosh2021hierarchy,Li2023AMHOTSC,Ghorashi2024AM,
Chatterjee2024,Bonetti2024HOTSC,Sun2025TOTSC,Liu2024SOTSC}, our system allows precise control over 
the number and spatial locations of MZMs simply by tailoring the edge termination, as shown in Figs.~\ref{fig5}(b) and \ref{fig5}(c). 
This tunability arises 
from a dimerized hopping structure in the Hamiltonian when both  $\eta_{1}$ and $\eta_{2}$ are non-zero, 
which makes the boundary topology highly sensitive to the sublattice (or layer) termination~\cite{supplemental,Teo2010}. 
Although such sublattice-dependent MZMs were previously predicted in heterostructures combining topological insulators and superconductors~\cite{Zhu2022sublattice,Zhu2023sublattice,Majid2022kagome}, here we 
demonstrate for the first time that this phenomenon can be realized in an intrinsic second-order TSC.

{\it Discussions and conclusions.---}The connection between opposite-sign pairing on DFPs and 
second-order topology was previously noted by Qin {\it et al.}~\cite{Qin2022hosc}, 
whose scenario involves two concentric DFPs and an extended $s$-wave pairing with nodes 
between them. Here, starting from an interacting Hamiltonian, we show that  
the superconducting ground state with  $d_{x^{2}-y^{2}}$-wave order parameter also leads to the same physics, 
even though the DFPs in our case enclose distinct Dirac points. 
This suggests the generality of the underlying 
principle and allows us to formulate a universal topological criterion: 
$(-1)^{\nu}=\prod_{n}\text{sgn}(\Delta)_{n}$, where $\nu=1$ ($0$) denotes the 
topological (trivial) phase and $\text{sgn}(\Delta)_{n}$ is the pairing sign 
on the $n$th DFP. 
Furthermore, we identify a novel tunability: the MZMs in our system can be repositioned 
by modifying the boundary termination, a feature not explored in the prior work.

To conclude, we have shown that intrinsic topological superconductivity can arise 
in DSMs from interaction-driven, even-parity spin-singlet pairing. The established 
mechanism can be directly generalized to three-dimensional 
nonsymmorphic DSMs~\cite{young2012dirac,Steinberg2014DSM,Gibson2015DSM,Wieder2016DSM}. Drawing parallels with
high-$T_{c}$ cuprates, our proposed $d$-wave second-order TSC may potentially be 
stabilized at elevated temperatures, providing a promising platform for MZMs that benefits 
from a large pairing gap and low disorder. Furthermore, the unique combination of 
$d$-wave pairing and Dirac Fermi surfaces in this phase should give rise to additional 
novel physics; for instance, the vortex properties are anticipated to differ 
significantly from those in conventional gapless $d$-wave superconductors~\cite{Soininen1994dwave,Ichioka1996dwave,Franz1996}.
Given the abundance of DSM materials~\cite{Young2017DSM,Guan2017DSM,Li2018DSM,Sato2018NSDSM,Jin2020DSM,Meng2022DSM,Chen2022DSM}, our simple topological criterion provides 
a clear guideline for screening candidates with the requisite Fermi surface 
configuration and superconductivity. Lastly, cold-atom systems, with their 
exceptional tunability, present an ideal alternative platform
for implementing our proposed scenario.

{\it Acknowledgements.---}This work is supported by the National Natural Science Foundation of China (Grant No.
12174455, No. 12474264, No. 12174453, No. 12574153), Guangdong Basic and Applied
Basic Research Foundation (Grant No. 2023B1515040023),
Guangdong Provincial Quantum Science Strategic Initiative (Grant No. GDZX2404007) and National Key R\&D Program
of China (Grant No. 2022YFA1404103).

\bibliography{dirac}

\begin{widetext}
\clearpage
\begin{center}
\textbf{\large Supplemental Material for ``Intrinsic Second-Order Topological Superconductors with Tunable Majorana Zero Modes''}\\
\vspace{4mm}
{Xiao-Jiao Wang$^{1,2,3}$, Yijie Mo$^{1,2,3}$, Zhi Wang$^{1,2}$, Zhigang Wu$^{4,*}$, Zhongbo Yan$^{1,2,3,\dag}$}\\
\vspace{2mm}
{\em $^1$School of Physics, Sun Yat-sen University, Guangzhou 510275, China}\\
{\em $^2$Guangdong Provincial Key Laboratory of Magnetoelectric Physics and Devices, Sun Yat-sen University, Guangzhou 510275, China}\\
{\em $^3$State Key Laboratory of Optoelectronic Materials and Technologies, Sun Yat-sen University, Guangzhou 510275, China}\\
{\em $^4$Quantum Science Center of Guangdong-Hong Kong-Macao Greater Bay Area (Guangdong), Shenzhen 508045, China}
\end{center}

\setcounter{equation}{0}
\setcounter{figure}{0}
\setcounter{table}{0}
\makeatletter
\renewcommand{\theequation}{S\arabic{equation}}
\renewcommand{\thefigure}{S\arabic{figure}}
\renewcommand{\bibnumfmt}[1]{[S#1]}

The supplemental material provides both bulk and boundary theories to understand the second-order topology and tunable Majorana zero modes (MZMs).
Three sections are in order: (I) Topological configuration in momentum space and second-order topology; (II)  Bulk topological invariant; (III) Edge-state theory for tunable MZMs.

%==================================================

\section{I.  Topological configuration in momentum space and second-order topology}

The principle of adiabatic continuity provides a powerful framework for analyzing the 
topological equivalence of gapped quantum phases. In essence, the principle asserts that 
two gapped phases are topologically equivalent if their respective Hamiltonians can be 
connected through a continuous path of parameters that never closes the energy gap and 
preserves the system's symmetries. The utility of this approach stems from the fact 
that a topologically trivial limit is often known for a given Hamiltonian. Consequently, 
one can determine whether a gapped phase is topological by checking if it can be adiabatically 
connected to this trivial limit. If such a connection is possible, the phase is trivial;
otherwise, it is topological.

For a Bogoliubov-de Gennes (BdG) Hamiltonian,  the limit $\mu\rightarrow\pm \infty$ is always topologically trivial, 
as it corresponds to a state where all bands are either completely filled  ($\mu=+\infty$) or completely empty ($\mu=-\infty$). 
Combining this fact with the principle of adiabatic continuity hence provides an efficient approach to determine 
whether a gapped superconducting state is topological or trivial. In the following, we apply this simple method 
to demonstrate that a superconducting state with two Dirac Fermi pockets gapped by opposite-sign pairing
is necessarily  topological. 

We start from the normal-state Hamiltonian, which reads 
\begin{eqnarray}
\mathcal{H}(\bk)=(\epsilon(\bk)-\mu)+\eta(\bk)\sigma_{x}+2\lambda_{\rm so}\sigma_{z}(\sin k_x s_y-\sin k_y s_x).
\end{eqnarray}
The functions $\epsilon(\bk) = -2t(\cos k_x + \cos k_y)$ and 
$\eta(\bk) = 4[\eta_1 \cos (k_x/2) \cos (k_y/2) + \eta_2 \sin (k_x/2) \sin (k_y/2)]$ are defined for brevity.
Here we have added the chemical potential $\mu$ to the Hamiltonian. Accordingly, the Fermi surface 
corresponds to the zero-energy contours of the energy spectrum. The explicit form of the energy spectrum is 
\begin{eqnarray}
E_{\pm}(\bk)=(\epsilon(\bk)-\mu)\pm\sqrt{\eta^{2}(\bk)+4\lambda_{\rm so}^{2}(\sin^{2}k_{x}+\sin^{2}k_{y})}.
\end{eqnarray}
Each band is doubly degenerate due to the conversation of $\mathcal{PT}$ symmetry. In Fig.~\ref{figS1}, we show 
the evolution of the Fermi surface with respect to $\mu$, while keeping all other parameters fixed.
At $\mu=0$, the Fermi level crosses the two Dirac points located at $\textbf{X}=(\pi,0)$ and $\textbf{Y}=(0,\pi)$, 
giving rise to two point-like Fermi surfaces. When $\mu$ is varied away from zero---either increased 
or decreased---two finite-sized Fermi pockets emerge, one enclosing $\textbf{X}$ and the other $\textbf{Y}$.
The two Fermi pockets will merge when $|\mu|$ is increased to a critical value, resulting in a Lifshitz transition 
of the Fermi surface. With a further increase of $|\mu|$, the Fermi surface will become a circle 
enclosing either $\boldsymbol{\Gamma}$ or $\textbf{M}$. Since the band structure lacks Dirac points at these high-symmetry points, the Fermi surface shrinks continuously to a point and vanishes as $|\mu|$ is increased further.

Now we move to the superconducting state. Since the $A_{g}$ pairing is favored, we focus on 
the BdG Hamiltonian corresponding to this pairing channel. Its general form is 
\begin{eqnarray}
\mathcal{H}_{\rm BdG}(\bk)=(\epsilon(\bk)-\mu)\tau_{z}+\eta(\bk)\tau_{z}\sigma_{x}
+2\lambda_{\rm so}(\sin k_x \tau_{z}\sigma_{z}s_y-\sin k_y \sigma_{z}s_x)
+\Delta(\bk)\tau_{y}s_{y}. 
\end{eqnarray}
The pairing term anticommutes with all other terms in the BdG Hamiltonian, and thus the BdG spectrum is simply given by 
\begin{eqnarray}
E_{\pm,\pm}(\bk)=\pm\sqrt{\left[(\epsilon(\bk)-\mu)\pm\sqrt{\eta^{2}(\bk)+4\lambda_{\rm so}^{2}(\sin^{2}k_{x}+\sin^{2}k_{y})}\right]^{2}+\Delta^{2}(\bk)}.
\end{eqnarray}
It is readily seen that the energy gap closes only when the pairing node (momentums satisfying $\Delta(\bk)=0$) meets 
the Fermi surface. When the Fermi surface contains two Dirac Fermi pockets near $\textbf{X}$ and $\textbf{Y}$, we have demonstrated that the pairing function takes the $d_{x^{2}-y^{2}}$-wave form [$\Delta(\bk)=\Delta(\cos k_{x}-\cos k_{y})$], resulting in a gapped superconducting state 
with opposite pairing signs on the two Fermi pockets. For 
this $d_{x^{2}-y^{2}}$-wave function, the pairing nodes lie along the lines  $k_{x}=k_{y}$ and $k_{x}=-k_{y}$, as 
indicated by the two dashed lines in Fig.~\ref{figS1}. 
With the pairing nodes fixed, Fig.~\ref{figS1} clearly shows that the Fermi 
pockets enclosing the Dirac points cannot be adiabatically deformed to vanish without 
crossing the pairing node lines. This obstruction intuitively demonstrates that 
this gapped $d$-wave superconducting phase is topologically distinct from the trivial phase at $\mu=\pm\infty$
and must therefore be topological. 

\begin{figure}[t]
\centering
\includegraphics[width=0.95\textwidth]{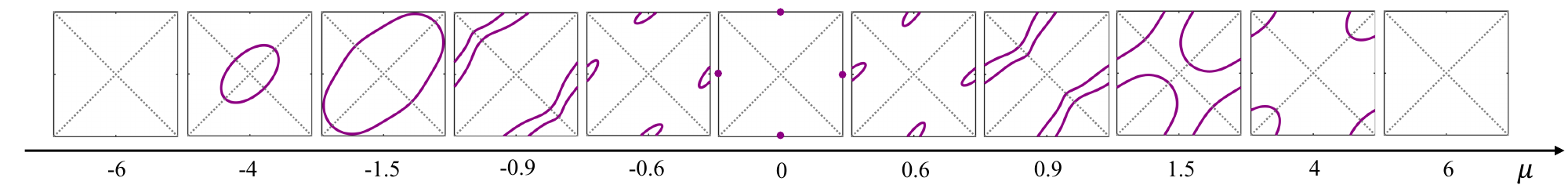}
\caption{Evolution of the Fermi surface (purple solid lines) in the first Brillouin zone 
with respect to the chemical potential $\mu$. 
Starting from a configuration with two Dirac Fermi points ($\mu=0$) separated 
by the pairing node lines (dashed lines), the Fermi surface cannot be made to vanish without crossing the pairing node lines. 
This indicates a topological configuration, as its adiabatic path to the trivial limit (without Fermi surface) is obstructed.
Common parameters: $t=0.5$, $\lambda_{\rm so}=0.4$, $\eta_{1}=0.8$, and $\eta_{2}=0.8$. 
}\label{figS1}
\end{figure}

\section{II.  Bulk topological invariant}\label{SecII}

Having intuitively established the topological nature of the superconducting state with two Dirac Fermi pockets gapped by a 
$d$-wave pairing, we now quantify its character. By applying the principle of adiabatic continuity, below we demonstrate that this phase is a second-order topological superconductor, characterized by a quantized quadrupole moment.

We note that the BdG Hamiltonian commutes with the mirror symmetry operator $\tilde{\mathcal{M}}_{z}=i\tau_{z}\sigma_{x}s_{z}$, i.e., 
$[\mathcal{H}_{\rm BdG}(\bk), \tilde{\mathcal{M}}_{z}]=0$. Therefore, it can be block-diagonalized into two sectors
according to the $\pm i$ eigenvalues of $\tilde{\mathcal{M}}_{z}$: $\mathcal{H}_{\rm BdG}=\mathcal{H}_{+i}\oplus \mathcal{H}_{-i}$. 
As these two mirror sectors are related by time-reversal symmetry, we can characterize the topology by focusing on a single sector. 
Without loss of generality, we choose $\mathcal{H}_{+i}$, whose  explicit form is 
\begin{eqnarray}
\mathcal{H}_{+i}(\bk)&=&-(\Delta(\bk)+\Delta_0)\tau_{x}\rho_{z}-2\lambda_{\text{so}}(\sin k_x\rho_y-\sin k_y\rho_x)\nonumber\\
&&+(\epsilon(\bk)-\mu)\tau_{z}+\eta(\bk)\tau_{z}\rho_{z}. \label{MH1}
\end{eqnarray}
To demonstrate the bulk-corner correspondence--the power of the bulk topological 
invariant to predict zero-energy corner states--we introduce an on-site
$s$-wave component $\Delta_{0}$ to the pairing function. This allows us to 
drive a direct transition between a second-order topological phase and a 
trivial phase simply by varying $\Delta_{0}$, while preserving all relevant symmetries.

In Fig.~\ref{figS1}, the three middle configurations are topologically equivalent, 
as they can be adiabatically connected without closing the energy gap. This allows us to simplify the analysis by setting 
$\mu=0$. Furthermore, at $\mu=0$, the Fermi surfaces are point-like and unaffected by $\epsilon(\bk)$.
We can therefore also set  $t=0$, thereby eliminating the $\epsilon(\bk)$ term from the Hamiltonian entirely. 

When $t=0$ and $\mu=0$, the mirror Hamiltonian $\mathcal{H}_{+i}$ reduces to
\begin{eqnarray}
\mathcal{H}_{+i}(\bk)=-(\Delta(\bk)+\Delta_0)\tau_{x}\rho_{z}-2\lambda_{\text{so}}(\sin k_x\rho_y-\sin k_y\rho_x)+\eta(\bk)\tau_{z}\rho_{z}. 
\end{eqnarray}
We perform a unitary transformation to the Hamiltonian: $\mathcal{H}^{\prime}_{+i}=W^{\dag} \mathcal{H}_{+i} W$ with $W=e^{i\frac{\pi}{4}\tau_{y}}$,
which yields
\begin{eqnarray}
\mathcal{H}^{\prime}_{+i}(\bk)=(\Delta(\bk)+\Delta_0)\tau_{z}\rho_{z}-2\lambda_{\text{so}}(\sin k_x\rho_y-\sin k_y\rho_x)+\eta(\bk)\tau_{x}\rho_{z}.
\label{MH2}
\end{eqnarray}
This form of the Hamiltonian admits a straightforward real-space interpretation, enabling the definition 
of a real-space topological invariant. Since a topological invariant is a mathematical property of the 
Hamiltonian itself and is independent of how we interpret the Pauli matrices, we can freely assign them 
for computational convenience. We therefore assign the $\tau_{i}$ matrices to act on the sublattice degrees 
of freedom and the $\rho_{i}$ matrices to act on an internal degree of freedom (such as spin).

It is readily seen that $\mathcal{H}^{\prime}_{+i}$ possesses chiral symmetry, with the symmetry operator given by 
$\mathcal{S} = \tau_y \rho_z$ and satisfying $\{\mathcal{H}^{\prime}_{+i}, \mathcal{S}\}=0$. 
This symmetry can quantize the quadrupole moment defined as~\cite{Yang2021HOTI,Kang2019multipole}:
\begin{eqnarray}
	q_{xy}&=&\left[  \frac {1}{2\pi}\text{Im} \log[\det (U^\dag\hat Q U)]-q_0\right]  \mod 1, \label{qxy}
\end{eqnarray}
where $\hat Q=\text{diag}\left\lbrace e^{2\pi i \hat x_j\hat y_j/(L_xL_y)}\right\rbrace^{4L_xL_y}_{j=1} $ with $\hat x_j(\hat y_j)$ denoting the $x$-position ($y$-position) operator for electron $j$, and the matrix $U$ is constructed by column-wise packing all the occupied eigenstates of an $L_x\times L_y$ system under periodic boundary conditions. Here, $q_0=\frac {1}{4\pi} \text{Im}\log \det \hat Q$ is the contribution from the background positive charge distribution. When \(q_{xy}=1/2\), the gapped phase is a second-order topological phase characterized 
by the presence of zero-energy corner states when open boundary conditions are imposed along both $x$ and $y$.

The quantization of $q_{xy}$ can be proved as follows. First, one can rewrite Eq. (\ref{qxy}) as
\begin{eqnarray}
q_{x y} & =&\left[\frac{1}{2 \pi} \operatorname{lm} \log \left[\operatorname{det}\left(U^{\dag } \hat{Q} U\right)\right]-q_{0}\right] \bmod 1  \nonumber\\
&=&\left[\frac{1}{2 \pi} \operatorname{lm} \log \left[\operatorname{det}\left(U^{\dag } \hat{Q} U\right)\right]-\frac{1}{4 \pi} \operatorname{Im} \log \operatorname{det} \hat{Q}\right] \bmod 1 \nonumber
\\&=&\left[\frac{1}{2 \pi} \operatorname{Im} \log \left[\operatorname{det}\left(U^{\dag} \hat{Q} U\right)\right]+\frac{1}{2 \pi} \operatorname{Im} \log \left(\operatorname{det} \hat{Q}^{\dag}\right)^{\frac{1}{2}}\right] \bmod 1 \nonumber
\\&=&\left[\frac{1}{2 \pi} \operatorname{Im} \log \left[\operatorname{det}\left(U^{\dag} \hat{Q} U\right)\right] \sqrt{\operatorname{det} \hat{Q}^{\dag}} \right]\bmod 1 .\label{qxy1}
\end{eqnarray}
Proving the quantization of the quadrupole moment is therefore equivalent to showing that chiral symmetry ensures the quantity $\det(U^\dagger \hat{Q} U) \sqrt{\det(\hat{Q}^\dagger)}$ is real~\cite{Li2020quadrupole}. To proceed, we deform the determinant as follows:
\begin{eqnarray}
\det (U^\dag\hat Q U)&=&\det [U^\dag(\hat Q-\boldsymbol{1}+\boldsymbol{1})U]\nonumber\\
&=&\det [\boldsymbol{1}+U^\dag(\hat Q-\boldsymbol{1})U].
\end{eqnarray}
Applying Sylvester's determinant identity, \(\det(\boldsymbol{1} + AB) = \det(\boldsymbol{1} + BA)\), simplifies the expression to
\begin{eqnarray}
	\det(U^{\dag} \hat{Q} U) = \det(\boldsymbol{1} + (\hat{Q} - \boldsymbol{1} )UU^{\dag}).
\end{eqnarray}
Recognizing that \(U U^\dagger = P_\text{occ}\) is the projection operator onto the occupied states, 
and that it can also be expressed as \(P_\text{occ} = \boldsymbol{1} - V V^\dagger\) (where \(V\) is 
the matrix of unoccupied state eigenvectors), we derive the following relation:
\begin{eqnarray}
	\det(U^{\dag}\hat{Q}U) &=& \det[\boldsymbol{1} + (\hat{Q} - \boldsymbol{1})(\boldsymbol{1} - VV^{\dag})] \nonumber\\
	&=& \det[\hat{Q} - (\hat{Q} - \boldsymbol{1} )VV^{\dag}] \nonumber\\
	&=& \det[\boldsymbol{1} + (\hat{Q}^{\dag} - \boldsymbol{1})VV^{\dag}]\det\hat{Q} \nonumber\\
	&=& \det(V^{\dag}\hat{Q}^{\dag}V)\det\hat{Q}.
\end{eqnarray}

Now, we invoke chiral symmetry, defined by \(\mathcal{S} \mathcal{H} \mathcal{S}^{-1} = -\mathcal{H}\). 
This symmetry implies a relation between the occupied and unoccupied eigenvector matrices:
\begin{eqnarray}
	V = \mathcal{S} U.
\end{eqnarray}
Using this relation, along with the commutation \([\mathcal{S}, \hat{Q}] = 0\) (note both matrices are diagonal), we can find
\begin{eqnarray}
	\det(U^{\dag}\hat{Q}U) &=& \det(V^{\dag}\hat{Q}^{\dag}V)\det\hat{Q} \nonumber\\
	&=& \det(U^{\dag}S^{\dag}\hat{Q}^{\dag}SU)\det\hat{Q} \nonumber\\
	&=& \det(U^{\dag}\hat{Q}^{\dag}U)\det\hat{Q}. \label{relation}
\end{eqnarray}
Since \(\hat{Q}\) is unitary, we have \(\det(\hat{Q}) \det(\hat{Q}^\dagger) = \det(\hat{Q} \hat{Q}^\dagger) = 1\). Combining this with 
Eq.~(\ref{relation}) yields:
\begin{eqnarray}
	\det(U^{\dag}\hat{Q}U)\sqrt{\det\hat{Q}^{\dag}} 
	= \det(U^{\dag}\hat{Q}^{\dag}U)\sqrt{\det\hat{Q}}.
\end{eqnarray}
Furthermore, using the property \(\det(\hat{Q}) = [\det(\hat{Q}^\dagger)]^*\), we find:
\begin{eqnarray}
	\det(U^{\dag}\hat{Q}U)\sqrt{\det\hat{Q}^{\dag}}
	= \left( \det(U^{\dag}\hat{Q}U)\sqrt{\det\hat{Q}^{\dag}} \right)^{*}. 
\end{eqnarray}
This equality confirms that \(\det(U^\dagger \hat{Q} U) \sqrt{\det\hat{Q}^\dagger}\) is real. 
Consequently, the quadrupole moment \(q_{xy}\) is quantized to \(0\) or \(1/2\), modulo 1.

\begin{figure}[t]
\centering
\includegraphics[width=0.8\textwidth]{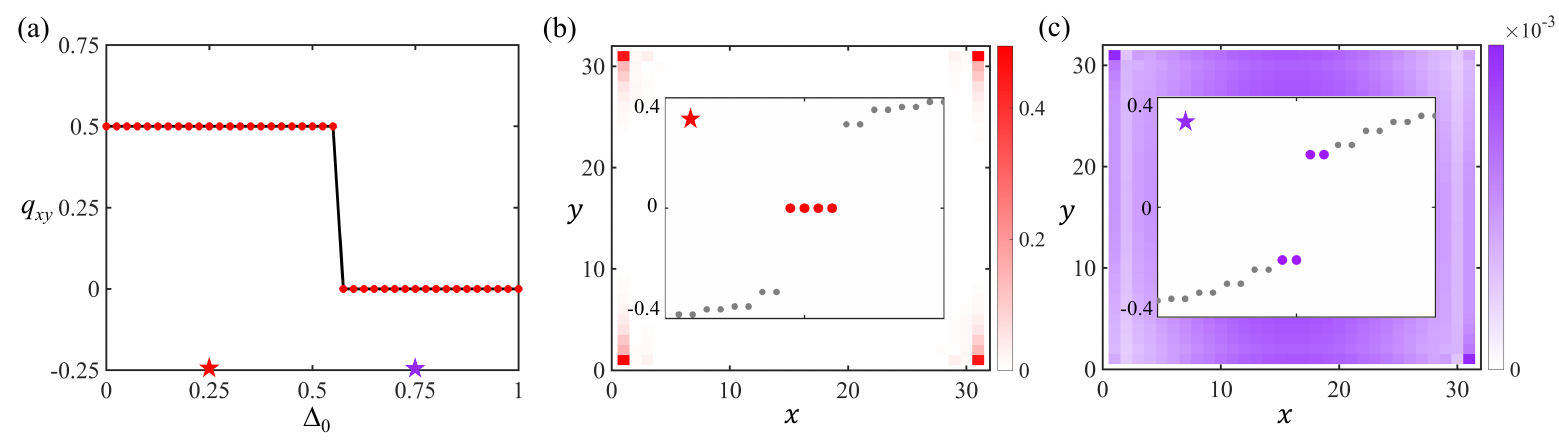}
\caption{ (a) Evolution of the quadrupole moment \(q_{xy} \) with respect to \( \Delta_0 \). (b) Four zero-energy states 
are found in the spectrum when $\Delta_{0}=0.25$. Their wave functions are localized at the four corners of the full open-boundary 
system. (c) No zero-energy states are found when  $\Delta_{0}=0.75$. Common parameters: $t=0$, $\mu=0$, $\lambda_{\rm so}=0.4$, 
$\eta_{1}=0.8$, $\eta_{2}=0.2$, and $\Delta=0.3$.  
}\label{figS2}
\end{figure}

Figure~\ref{figS2}(a) shows the evolution of the quadrupole moment 
\(q_{xy} \) with respect to \( \Delta_0 \). The sharp change at  \( \Delta_0 \approx 0.55 \) 
indicates a phase transition from a topological to a trivial gapped phase. 
To verify the bulk-corner correspondence, we compute the energy spectrum 
under open boundary conditions. For a representative topological phase with
$q_{xy} = 1/2$, the spectrum in Fig.~\ref{figS2}(b) reveals four zero-energy modes, 
with wavefunctions localized at the corners. This confirms a second-order 
topological phase. In contrast, for a trivial phase with 
$q_{xy} =0$, Fig.~\ref{figS2}(c) shows no in-gap states, 
confirming the bulk's trivial nature.

\section{III.  Edge-state theory for tunable MZMs}

Although the bulk topological invariant predicts the existence of MZMs 
at the corners under full open boundary conditions, it cannot explain the sensitive 
dependence of their positions on the sublattice 
termination. Here, `sublattice' denotes the layer degree of freedom, which acts as a sublattice 
from a top-down view of the bilayer system.
To explain this intriguing phenomenon, 
we develop an edge-state theory, which provides a natural framework for fully understanding the behavior 
of topological boundary states. 

To derive the edge-state theory, we begin with the mirror-block Hamiltonians:
\begin{eqnarray}
\mathcal{H}_{\pm i}(\bk)&=&(\epsilon(\bk)-\mu)\tau_{z}+\eta(\bk)\tau_{z}\rho_{z}\mp \Delta(\bk)\tau_{x}\rho_{z}\nonumber\\
&&+2\lambda_{\rm so}(\mp\sin k_{x}\rho_{y}+\sin k_{y}\rho_{x}).
\end{eqnarray}
Since the analysis for the two mirror sectors is identical, we focus on $\mathcal{H}_{+i}(\mathbf{k})$. 
We decompose this Hamiltonian into two parts: $\mathcal{H}_{+ i}(\mathbf{k}) = \mathcal{H}_{0}(\mathbf{k}) 
+ \mathcal{H}_{p}(\mathbf{k})$, where
\begin{eqnarray}
\mathcal{H}_{0}(\bk)&=&-\Delta(\bk)\tau_{x}\rho_{z}-2\lambda_{\rm so}(\sin k_{x}\rho_{y}-\sin k_{y}\rho_{x}),\nonumber\\
\mathcal{H}_{p}(\bk)&=&(\epsilon(\bk)-\mu)\tau_{z}+\eta(\bk)\tau_{z}\rho_{z}.
\end{eqnarray}
We will treat $\mathcal{H}_{p}$ as a perturbation. Although the energy scale of $\mathcal{H}_{p}$ in realistic materials may be larger than that of $\mathcal{H}_{0}$, this approach is formally justified due to the principle of adiabatic continuity. Specifically,   
we can introduce a small, dimensionless parameter $\alpha$ to continuously tune the perturbation to $\alpha\mathcal{H}_{p}(\mathbf{k})$. 
This deformation preserves the system's symmetries and, crucially, does not close the bulk energy gap, thereby leaving the topology invariant. 
Because the topological physics is qualitatively unchanged under the adiabatic connection to $\alpha\mathcal{H}_{p}$ for small $\alpha$, we treat $\mathcal{H}_{p}$ directly as a perturbation.

Since the system has a bipartite-lattice structure, there are two natural unit cell choices. Interestingly, 
they lead to distinct sublattice terminations at the same boundary,  as shown in Figs.~\ref{figS3}(a) and \ref{figS3}(b). 
Below, we demonstrate that this termination profoundly affects the boundary topology.

While directly solving for 
the edge-state wave functions of this lattice Hamiltonian is achievable, focusing on the low-energy regime 
simplifies the analysis of the boundary topology. To establish a low-energy theory, we again invoke the 
principle of adiabatic continuity. Specifically, we consider a deviation from the perfect $d$-wave pairing 
symmetry to the form $\Delta(\mathbf{k})=\Delta\cos k_{x}-(\Delta+\delta)\cos k_{y}$, where both $\Delta$ 
and $\delta$ are positive constants and $\delta$ is assumed to be much smaller than all other energy scales. 
The introduction of this small parameter obviously affects neither the system's underlying symmetries nor 
closes the bulk energy gap; therefore, it leaves the topology of the full Hamiltonian unchanged. However, 
this small deviation opens a finite gap in the spectrum of $\mathcal{H}_{0}$, driving it into a weak topological 
insulator phase with helical edge states on the $y$-normal edges. Since $\mathcal{H}_{0}$ features two degenerate 
energy minima at $(0,0)$ and $(\pi,\pi)$, we perform a low-energy expansion around these two momenta, which yields:
\begin{eqnarray}
\mathcal{H}_{0}(\bq)_{(0,0)}&=&[\delta+\frac{\Delta}{2}q_{x}^{2}-\frac{1}{2}(\Delta+\delta) q_{y}^{2}]\tau_{x}\rho_{z}
-2\lambda_{\rm so}(q_{x}\rho_{y}-q_{y}\rho_{x}),\nonumber\\
\mathcal{H}_{0}(\bq)_{(\pi,\pi)}&=&-[\delta+\frac{\Delta}{2}q_{x}^{2}-\frac{1}{2}(\Delta+\delta) q_{y}^{2}]\tau_{x}\rho_{z}
+2\lambda_{\rm so}(q_{x}\rho_{y}-q_{y}\rho_{x}).
\end{eqnarray}

\begin{figure}[t]
\centering
\includegraphics[width=0.8\textwidth]{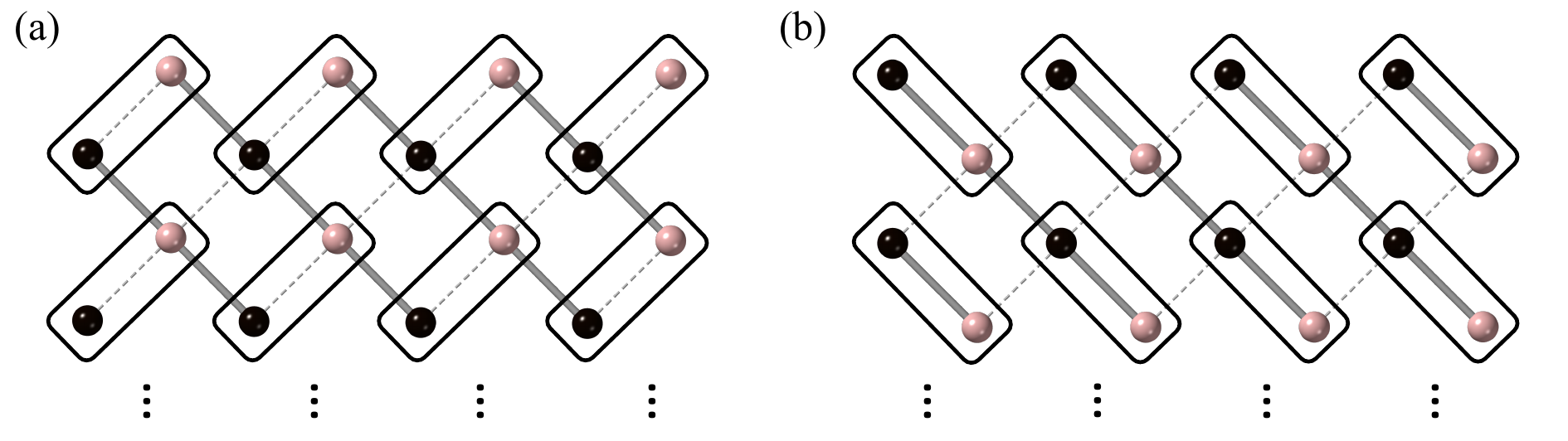}
\caption{Two distinct unit cell choices (rectangles) result in different sublattice 
terminations under open boundary conditions. (a) Type I: The unit cell 
leads to an upper edge with B-sublattice termination (pink dots). The sequence of sublattice 
is B-A-...-B-A from top to bottom. (b) Type II: The alternative unit cell  
gives A-sublattice termination (black dots) with sublattice sequence A-B-...-A-B. 
Bond conventions: solid lines for $\eta_{+}=\eta_{1}+\eta_{2}$, dashed lines for $\eta_{-}=\eta_{1}-\eta_{2}$.
}\label{figS3}
\end{figure}

In the main text, we employ a Fourier transform convention that explicitly specifies the sublattice positions:
\begin{eqnarray}
c_{\bR_{n},A}&=&\sum_{\bk}c_{\bk,A}e^{i\bk\cdot(\bR_{n}+\boldsymbol{\delta}_{A})},\nonumber\\
c_{\bR_{n},B}&=&\sum_{\bk}c_{\bk,B}e^{i\bk\cdot(\bR_{n}+\boldsymbol{\delta}_{B})},
\end{eqnarray}
where $\mathbf{R}_{n}$ is the position of the $n$th unit cell, and $\boldsymbol{\delta}_{A/B}$ 
are the basis vectors for the two sublattices. In this convention, the $\eta(\mathbf{k})$ term in 
the normal-state Hamiltonian takes the form
\begin{eqnarray}
\mathcal{H}_{\eta}(\bk)=\left(
                          \begin{array}{cccc}
                            0 & 0 & \eta(\bk) & 0 \\
                            0 & 0 & 0 & \eta(\bk) \\
                            \eta(\bk) & 0 & 0 & 0 \\
                            0 & \eta(\bk) & 0 & 0 \\
                          \end{array}
                        \right)=\eta(\bk)\sigma_{x}s_{0}.
\end{eqnarray} 
The function $\eta(\mathbf{k}) = 4(\eta_{1} \cos (k_{x}/2) \cos (k_{y}/2) + \eta_{2} \sin (k_{x}/2) \sin (k_{y}/2))$ 
has the property $\eta(\mathbf{k}+\mathbf{G}_{x}) = \eta(\mathbf{k}+\mathbf{G}_{y}) = -\eta(\mathbf{k})$, 
where $\mathbf{G}_{x}=(2\pi,0)$ and $\mathbf{G}_{y}=(0,2\pi)$ are reciprocal lattice vectors. Consequently, the full Hamiltonian is not periodic in the first Brillouin zone: $\mathcal{H}(\mathbf{k}+\mathbf{G}_{x/y}) \neq \mathcal{H}(\mathbf{k})$.
While the energy spectrum is independent of the Fourier transform convention, a periodic Hamiltonian is 
essential for a consistent topological analysis. We therefore adopt a convention that ensures $\mathcal{H}(\mathbf{k}+\mathbf{G}_{x}) = \mathcal{H}(\mathbf{k}+\mathbf{G}_{y}) = \mathcal{H}(\mathbf{k})$. This is achieved by defining the 
Fourier transform without the sublattice phase factors:
\begin{eqnarray}
c_{\bR_{n},A}&=&\sum_{\bk}c_{\bk,A}e^{i\bk\cdot\bR_{n}},\nonumber\\
c_{\bR_{n},B}&=&\sum_{\bk}c_{\bk,B}e^{i\bk\cdot\bR_{n}}.
\end{eqnarray}
With this convention, while intra-sublattice hopping and spin-orbit 
coupling terms retain their original form, the $\eta(\mathbf{k})$ term becomes 
dependent on the unit cell geometry. In the basis $\psi_{\mathbf{k}}=(c_{\mathbf{k},A\uparrow},c_{\mathbf{k},A\downarrow},c_{\mathbf{k},B\uparrow},c_{\mathbf{k},B\downarrow})^{T}$ 
and for the type-I unit cell shown in Fig.~\ref{figS3}(a), this term is given by
\begin{eqnarray}
\mathcal{H}_{\eta}^{(\text{I})}(\bk)=\left(
                          \begin{array}{cccc}
                            0 & 0 & g(\bk) & 0 \\
                            0 & 0 & 0 & g(\bk) \\
                            g^{*}(\bk) & 0 & 0 & 0 \\
                            0 & g^{*}(\bk) & 0 & 0 \\
                          \end{array}
                        \right)
\end{eqnarray} 
where 
\begin{eqnarray}
g(\bk)&=&\eta_{-}+\eta_{+}e^{-ik_{x}}+\eta_{+}e^{-ik_{y}}+\eta_{-}e^{-ik_{x}-ik_{y}}\\
&=&e^{-i\frac{(k_{x}+k_{y})}{2}}(4\eta_{1}\cos\frac{k_{x}}{2}\cos\frac{k_{y}}{2}
+4\eta_{2}\sin\frac{k_{x}}{2}\sin\frac{k_{y}}{2})\nonumber\\
&=&e^{-i\frac{(k_{x}+k_{y})}{2}}\eta(\bk).
\end{eqnarray}
It is evident that now the full Hamiltonian 
is periodic since $g(\mathbf{k}+\mathbf{G}_{x/y})=g(\bk)$. Also in the basis $\psi_{\mathbf{k}}=(c_{\mathbf{k},A\uparrow},c_{\mathbf{k},A\downarrow},c_{\mathbf{k},B\uparrow},c_{\mathbf{k},B\downarrow})^{T}$ and
for  the unit cell shown in Fig.~\ref{figS3}(b),  the form becomes 
\begin{eqnarray}
\mathcal{H}_{\eta}^{(\text{II})}(\bk)=\left(
                          \begin{array}{cccc}
                            0 & 0 & f(\bk) & 0 \\
                            0 & 0 & 0 & f(\bk) \\
                            f^{*}(\bk) & 0 & 0 & 0 \\
                            0 & f^{*}(\bk) & 0 & 0 \\
                          \end{array}
                        \right)
\end{eqnarray} 
where 
\begin{eqnarray}
f(\bk)&=&\eta_{+}+\eta_{-}e^{-ik_{x}}+\eta_{-}e^{ik_{y}}+\eta_{+}e^{-ik_{x}+ik_{y}}\nonumber\\
&=&e^{-i\frac{(k_{x}-k_{y})}{2}}(4\eta_{1}\cos\frac{k_{x}}{2}\cos\frac{k_{y}}{2}
+4\eta_{2}\sin\frac{k_{x}}{2}\sin\frac{k_{y}}{2})\nonumber\\
&=&e^{-i\frac{(k_{x}-k_{y})}{2}}\eta(\bk).
\end{eqnarray}
The two unit cell choices differ only by a phase factor. As shown below, 
this phase difference critically influences the boundary Dirac mass.

To construct a low-energy theory, we expand $\mathcal{H}_{\eta}^{(\text{I,II})}(\mathbf{k})$ to 
leading order around the two low-energy points, $(0,0)$ and $(\pi,\pi)$, obtaining:
\begin{eqnarray}
\mathcal{H}_{\eta}^{(\text{I})}(\bq)_{(0,0)}&=&4\eta_{1}\sigma_{x}s_{0},\nonumber\\
\mathcal{H}_{\eta}^{(\text{I})}(\bq)_{(\pi,\pi)}&=&-4\eta_{2}\sigma_{x}s_{0},\nonumber\\
\mathcal{H}_{\eta}^{(\text{II})}(\bq)_{(0,0)}&=&4\eta_{1}\sigma_{x}s_{0},\nonumber\\
\mathcal{H}_{\eta}^{(\text{II})}(\bq)_{(\pi,\pi)}&=&4\eta_{2}\sigma_{x}s_{0}.
\end{eqnarray}
Consequently, in the superconducting state, the perturbative Hamiltonians  (to leading order) for the two unit cell choices are:
\begin{eqnarray}
\mathcal{H}_{p}^{(\text{I})}(\bq)_{(0,0)}&=&(-4t-\mu)\tau_{z}+4\eta_{1}\tau_{z}\rho_{z},\nonumber\\
\mathcal{H}_{p}^{(\text{I})}(\bq)_{(\pi,\pi)}&=&(4t-\mu)\tau_{z}-4\eta_{2}\tau_{z}\rho_{z},\nonumber\\
\mathcal{H}_{p}^{(\text{II})}(\bq)_{(0,0)}&=&(-4t-\mu)\tau_{z}+4\eta_{1}\tau_{z}\rho_{z},\nonumber\\
\mathcal{H}_{p}^{(\text{II})}(\bq)_{(\pi,\pi)}&=&(4t-\mu)\tau_{z}+4\eta_{2}\tau_{z}\rho_{z}.
\end{eqnarray}
Since $\mathcal{H}_{0}$ consists entirely of intra-sublattice terms, it is identical 
for the two unit cell choices.

Next, we derive the low-energy edge-state Hamiltonian corresponding to each low-energy bulk Hamiltonian. 
We begin with 
\begin{eqnarray}
\mathcal{H}_{0}^{(\text{I})}(\bq)_{(0,0)}&=&[\delta+\frac{\Delta}{2}q_{x}^{2}-\frac{1}{2}(\Delta+\delta) q_{y}^{2}]\tau_{x}\rho_{z}
-2\lambda_{\rm so}(q_{x}\rho_{y}-q_{y}\rho_{x}),\nonumber\\
\mathcal{H}_{p}^{(\text{I})}(\bq)_{(0,0)}&=&(-4t-\mu)\tau_{z}+4\eta_{1}\tau_{z}\rho_{z}.
\end{eqnarray}
We now derive the edge states for the upper $y$-normal edge. For simplicity, consider the system 
in the half-infinite plane with $y \leq 0$, such that the upper edge corresponds to $y=0$. 
Solving the eigenvalue equation
$\mathcal{H}_{0}^{(\text{I})}(q_{x},q_{y}\rightarrow -i\partial_{y})_{(0,0)}\psi_{\alpha}(x,y)=E_{\alpha}\psi_{\alpha}(x,y)$ 
under the boundary conditions $\psi_{\alpha}(x,0) = \psi_{\alpha}(x,-\infty) = 0$ yields two branches of edge states. 
The first has a linear dispersion $E_1 = v q_x$ with velocity $v = 2\lambda_{\mathrm{so}}$, and a wave function 
given by~\cite{Yan2018hosc}
\begin{eqnarray}
\psi_{1}(x,y)=\mathcal{N}\sin(\kappa_{1}y) e^{\kappa_{2} y}e^{iq_{x}x}|\chi_{1}\rangle.
\end{eqnarray}
Here, $|\chi_{1}\rangle = |\tau_x = 1, \rho_y = -1\rangle$ is an eigenvector of $\tau_x \rho_y$, 
and $\mathcal{N} = 2\sqrt{ |\kappa_2 (\kappa_1^2 + \kappa_2^2) / \kappa_1^2 | }$ is the normalization constant, 
with $\kappa_1 = \sqrt{ (2\delta + \Delta q_x^2)/(\Delta + \delta) - v^2/(\Delta+\delta)^2 }$ and $\kappa_2 = v/(\Delta+\delta)$ (we have 
taken all coefficients to be positive for discussion convenience).
The second branch has the opposite chirality, with dispersion $E_2 = -v q_x$ and wave function
\begin{eqnarray}
\psi_{2}(x,y)=\mathcal{N}\sin(\kappa_{1}y) e^{\kappa_{2} y}e^{iq_{x}x}|\chi_{2}\rangle,
\end{eqnarray}
where $|\chi_{2}\rangle = |\tau_x = -1, \rho_y = 1\rangle$. Thus, before including perturbations, 
the low-energy edge-state Hamiltonian in the basis $(\psi_1, \psi_2)^T$ is
\begin{eqnarray}
\mathcal{H}^{(\text{I})}(q_{x})_{(0)}=vq_{x}\zeta_{z},
\end{eqnarray}
where $\zeta_i$ are Pauli matrices in the two-dimensional Hilbert space spanned by the edge states. Projecting the perturbation Hamiltonian $\mathcal{H}_{p}^{\mathrm{(I)}}$ onto this subspace gives
\begin{eqnarray}
\int_{-\infty}^{0}\psi_{\alpha}^{\dag}(x,y)\mathcal{H}_{p}^{(\text{I})}(q_{x},-i\partial_{y})_{(0,0)}\psi_{\beta}(x,y)dy=(4\eta_{1}\zeta_{x})_{\alpha\beta}.
\end{eqnarray}
Consequently, the full low-energy edge-state Hamiltonian for this case is
\begin{eqnarray}
\mathcal{H}^{(\text{I})}(q_{x})_{(0)}=vq_{x}\zeta_{z}+m_{1}\zeta_{x},
\end{eqnarray}
with Dirac mass $m_1 = 4\eta_1$. Notably, the $(-4t - \mu)\tau_z$ term does not contribute to 
the boundary Hamiltonian, indicating its irrelevance to the topological boundary physics in this regime. 
This finding provides an intuitive justification for neglecting this term when calculating the bulk topological invariant in Sec.~II.

The same procedure applies to the other three cases, yielding the low-energy edge-state Hamiltonians:
\begin{eqnarray}
\mathcal{H}^{(\text{I})}(q_{x})_{(\pi)}&=&-vq_{x}\zeta_{z}-m_{2}\zeta_{x},\nonumber\\
\mathcal{H}^{(\text{II})}(q_{x})_{(0)}&=&vq_{x}\zeta_{z}+m_{1}\zeta_{x},\nonumber\\
\mathcal{H}^{(\text{II})}(q_{x})_{(\pi)}&=&-vq_{x}\zeta_{z}+m_{2}\zeta_{x},
\end{eqnarray}
with Dirac mass $m_{2}=4\eta_{2}$ and the same basis $(\psi_1, \psi_2)^T$. 

Each low-energy edge-state Hamiltonian possesses chiral symmetry 
($\mathcal{S} = \zeta_y$), allowing characterization by a winding number:
\begin{eqnarray}
W_{\beta}^{(a)}=\frac{i}{4\pi}\int_{-\infty}^{\infty}dq_{x}\text{Tr}\{\mathcal{S}[\mathcal{H}^{(a)}(q_{x})_{(\beta)}]^{-1}
\partial_{q_{x}}\mathcal{H}^{(a)}(q_{x})_{(\beta)}\},
\end{eqnarray}
where $a \in \{\mathrm{I}, \mathrm{II}\}$ and $\beta \in \{0, \pi\}$. Assuming $\eta_1, \eta_2 > 0$, a direct calculation gives:
\begin{eqnarray}
&W_{0}^{(\text{I})}=\frac{1}{2}, W_{\pi}^{(\text{I})}=\frac{1}{2};\nonumber\\
&W_{0}^{(\text{II})}=\frac{1}{2}, W_{\pi}^{(\text{II})}=-\frac{1}{2}.
\end{eqnarray}
The winding number characterizing the full edge is $W_T^{(a)} = W_0^{(a)} + W_\pi^{(a)}$. For the type-I edge in Fig.~\ref{figS3}(a), we find:
\begin{eqnarray}
W_{T}^{(\text{I})}=W_{0}^{(\text{I})}+W_{\pi}^{(\text{I})}=1,
\end{eqnarray}
while for the type-II edge shown in Fig.~\ref{figS3}(b): 
\begin{eqnarray}
W_{T}^{(\text{II})}=W_{0}^{(\text{II})}+W_{\pi}^{(\text{II})}=0.
\end{eqnarray}

Since the only difference between the edges in Figs.~\ref{figS3}(a) and \ref{figS3}(b) is their outermost sublattice, 
the different values for $W_{T}^{(\text{I})}$ and $W_{T}^{(\text{II})}$ indicate that the boundary topology depends sensitively on the sublattice termination. Consequently, when an edge contains segments with different sublattice terminations (A versus B), their interface 
forms a \textit{sublattice domain wall}~\cite{Zhu2022sublattice,Zhu2023sublattice,Majid2022kagome}. Across this topological defect, the winding number changes by $\Delta W_T =|W_{T}^{(\text{I})}-W_{T}^{(\text{II})}|= 1$.
According to bulk-defect correspondence~\cite{Teo2010}, this winding number difference dictates the number of zero-energy 
bound states at the defect. With $\Delta W_T = 1$ per mirror sector, each sublattice domain wall binds two zero-energy states. 
This winding-number analysis based on low-energy theory explains why MZM positions in this system can 
be manipulated by engineering the sublattice termination.

\end{widetext}

\end{document}